\documentclass[12pt,preprint]{aastex}

\slugcomment{Accepted by ApJL}

\def\MO{M_\odot}
\def\LO{L_\odot}

\shorttitle{IRAS\,16547$-$4247: A New Candidate of a Protocluster Unveiled with ALMA}
\shortauthors{A. E. Higuchi et al.}

\begin{document}
\title{IRAS\,16547$-$4247: A New Candidate of a Protocluster Unveiled with ALMA}

\author{Aya E. Higuchi\altaffilmark{1}, Kazuya Saigo\altaffilmark{2}, James O. Chibueze\altaffilmark{2,3}, Patricio Sanhueza\altaffilmark{2},
Shigehisa Takakuwa\altaffilmark{4}, and Guido Garay\altaffilmark{5}}
\email{ahiguchi@mx.ibaraki.ac.jp}

\altaffiltext{1}{Ibaraki University, College of Science, 2-1-1 Bunkyo, Mito, 310-8512, Japan}
\altaffiltext{2}{National Astronomical Observatory of Japan 2-21-1 Osawa, Mitaka, Tokyo, 181-8588, Japan}
\altaffiltext{3}{Department of Physics and Astronomy, Faculty of Physical Sciences, University of Nigeria, Carver Building, 1 University Road, Nsukka, Nigeria.}
\altaffiltext{4}{Academia Sinica Institute of Astronomy and Astrophysics, P.O. Box 23-141, Taipei 10617, Taiwan}
\altaffiltext{5}{Departamento de Astronom\'\i a, Universidad de Chile, Camino El Observatorio 1515, Las Condes, Santiago, Chile}

\begin{abstract}

We present the results of continuum and $^{12}$CO(3--2) and CH$_{3}$OH(7--6) line 
observations of IRAS\,16547$-$4247 made with the Atacama Large 
Millimeter/submillimeter Array (ALMA) at an angular resolution of $\sim$0\farcs5. 
The $^{12}$CO(3--2) emission shows two high-velocity outflows whose driving sources are located within the dust continuum 
peak. The alignment of these outflows do not coincide with that of the wide-angle, 
large scale, bipolar outflow detected with APEX in previous studies. 
The CH$_{3}$OH(7--6) line emission traces an $\it{hourglass}$ structure 
associated with the cavity walls created by the outflow lobes.
Taking into account our results together with the position of the H$_{2}$O and class I CH$_{3}$OH maser clusters, 
we discuss two possible scenarios that can explain the hourglass structure observed in IRAS\,16547$-$4247:
(1) precession of a biconical jet, (2) multiple, or at least two, driving 
sources powering intersecting outflows. Combining the available evidence, 
namely, the presence of two cross-aligned bipolar outflows and two different 
H$_{2}$O maser groups, we suggest that IRAS\,16547$-$4247 represents an early 
formation phase of a protocluster.

\end{abstract}


\section{INTRODUCTION}

Most stars, particularly high-mass stars ($>$$\,$8$~\MO$), form in clusters 
(Lada $\&$ Lada 2003). Stellar clusters form in dense and massive molecular clumps 
(size$\sim$$\,$1~pc, mass$\sim$$\,$100--1000~$\MO$, density 
$\sim$10$^{4-5}$~$\rm{cm}^{-3}$) \citep{rid03, lad03, lad10, hig09, hig10, hig13}.
The gas in these clumps can be significantly affected by the feedback from newly 
formed stars, thus blurring our understanding of cluster formation. 
To explore the initial conditions and details of cluster formation,
it is then necessary to study molecular clouds in the early stages of evolution at high-angular resolution 
(e.g., Higuchi et al. 2014).
High-mass stars are usually deeply embedded in their parent cloud, obscuring their early formative stages. 
Their formation timescales of $\sim$ 10$^{5}$~yr are short, and they form in distant clusters and associations 
(e.g., McKee $\&$ Ostriker 2007; Zinnecker $\&$ Yorke 2007). 
These factors also contribute to our poor understanding of their formation processes.
High angular resolution observations are indispensable in the efforts to unveil the mystery of high-mass star formation. 
Millimeter and sub-millimeter interferometric observations have contributed largely to the current 
knowledge of high-mass star formation. 
The Atacama Large Millimeter/submillimeter Array (ALMA) provides the high sensitivity, angular 
resolution, and dynamic range to improve our understanding of the formation processes of high-mass stars.

IRAS\,16547$-$4247 is a luminous infrared source (bolometric luminosity of 
6.2~$\times$~10$^{4}$~$\LO$), located at a distance of $\sim$2.9~kpc \citep{rod08}.
Single-dish dust continuum observations show that the IRAS source 
corresponds to a dense molecular clump with a mass of 1.3 $\times$ 10$^{3}$~$\MO$ 
and a radius of 0.2 pc \citep{gar03}.
Very Large Array (VLA) radio continuum observations revealed the presence of a 
thermal radio jet located at the center of the core \citep{gar03, rod05}.
\cite{rod05} detected, in addition, several fainter radio sources suggesting that they are probably 
members of a young cluster.
\cite{gar07} identified a large scale bipolar $^{12}$CO($J$=3--2) outflow, which 
is roughly oriented in the north-south direction and centered at the position of the 
jet source. \cite{fra09} reported 1.3~mm dust continuum and SO$_{2}$ observations 
made with the Submillimeter Array (SMA). They found that the SO$_{2}$ emission 
traces a small molecular structure associated with the jet exhibiting 
a velocity gradient and proposed the existence of Keplerian rotation. High angular 
resolution observations of the high-velocity outflows towards IRAS\,16547$-$4247 have not, however, been carried out. 

In this paper, we present $\sim$0\farcs5 resolution images
of the dust continuum and $^{12}$CO($J$=3--2)[hereafter $^{12}$CO(3--2)] and 
CH$_{3}$OH($v_{t}$=0,$J_{K}$=7$_{0}$--6$_{0}$)[hereafter CH$_{3}$OH(7--6)] line emission using ALMA. 
The goal is to determine the spatial distribution and velocity 
structure of the gas around the center of IRAS\,16547$-$4247, in order to explore 
the dynamical processes associated with the information of high-mass stars formation. 

\section{OBSERVATIONS}

\begin{figure}
\epsscale{1}
\plotone{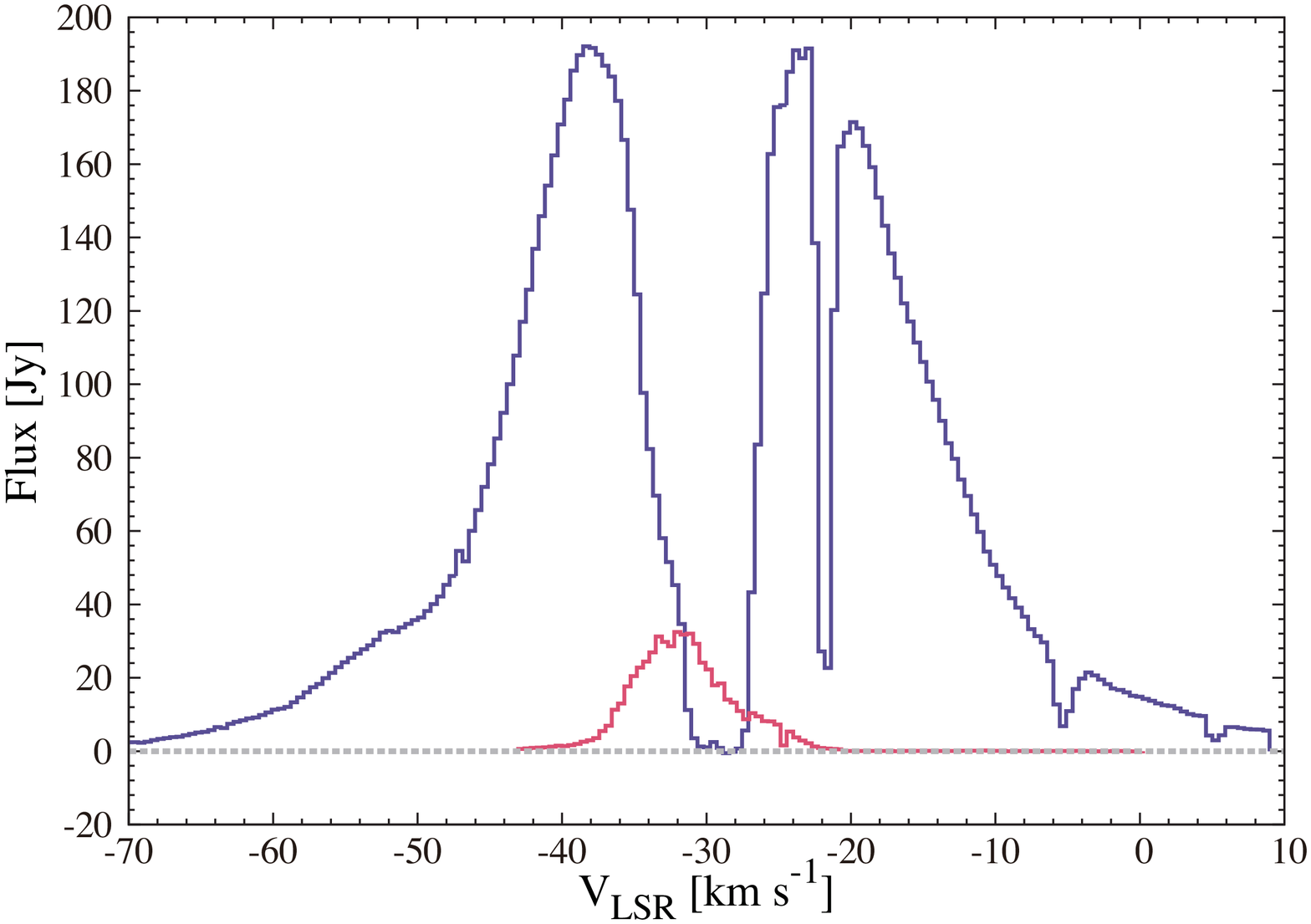}
\caption{
Average spectra of two lines detected from IRAS\,16547$-$4247: $^{12}$CO(3--2) (purple) and CH$_{3}$OH(7--6) (pink). 
The spectra were averaged over a region of 23$\arcsec$ centered on the continuum cores.
This region corresponds to a box with corners at ($\alpha_{J2000}$, $\delta_{J2000}$)= [16$^{\mbox h}$58$^{\mbox m}$18$^{\mbox s}.$3, 
~$-$~$\!\!$042$^{\circ}$52$^{\arcmin}$19$\arcsec.$62] and ($\alpha_{J2000}$, 
$\delta_{J2000}$)=[16$^{\mbox h}$58$^{\mbox m}$16$^{\mbox s}.$17, ~$-$~$\!\!$042$^{\circ}$51$^{\arcmin}$56$\arcsec.$72].}
\label{fig1}
\end{figure}

IRAS\,16547$-$4247 was observed with ALMA \citep{hil10} during Early Science Cycle 0 
in its extended configuration, using the band 7 receivers. The observations were 
done in 4 executions with 32~12-m antennas at an angular resolution of $\sim$0\farcs5.
At the distance of IRAS\,16547$-$4247 (d$\sim$2.9~kpc), the angular size corresponds to 0.007~pc.
In this work, we present results of the 880~$\mu$m dust continuum emission, $^{12}$CO(3--2) [frequency of 345.796~GHz]
and CH$_{3}$OH(7--6) [frequency of 338.409~GHz] lines (see Figure \ref{fig1}).

The ALMA calibration includes simultaneous observations of the 183~GHz water line 
with water vapor radiometers that measure the water column in the antenna beam, 
later used to reduce the atmospheric phase noise. Amplitude calibration was done using Titan.
The quasars 3c279 and J1604-446 were used to calibrate the bandpass 
and the complex gain fluctuations, respectively. The continuum map was obtained 
from a combination of all the line-free channels. 
Briggs weighting with a robust parameter of $+$0.5 was used in both continuum and line images. 
Data reduction was performed using version 3.4 of the Common Astronomy Software Applications package 
(http://casa.nrao.edu).
The CASA task $\it{CLEAN}$ was used to Fourier-transform the visibility data and deconvolve the dirty images,
at an velocity interval of 0.44~km~s$^{-1}$.

\section{RESULTS}

\subsection{880~$\mu$m continuum emission}

Figure \ref{fig2}(a) shows the ALMA 880~$\micron$ dust continuum map at a 0\farcs5 resolution. 
We resolved the SMA submillimeter source reported by \cite{fra09} into 
two sources separated by $\sim$2$\arcsec$ (core-A and core-B in Figure \ref{fig2}(a)).
By using the CASA task $\it{imfit}$, a Gaussian fit to the continuum emission was performed.
We obtain that the peak flux density at 880~$\micron$ is 420$\pm$30~mJy~beam$^{-1}$ for core-A and 100$\pm$10~mJy~beam$^{-1}$ for core-B.
The deconvolved sizes are 1\farcs1$\times$0\farcs7 with a P.A.=127$^{\circ}$ (integrated flux: 2.2~Jy) for core-A 
and 2\farcs8$\times$1\farcs0 with a P.A.=69$^{\circ}$ (integrated flux: 1.7~Jy) for core-B. 
The peak position of core-A, ($\alpha_{J2000}$, $\delta_{J2000}$)=16$^{\mbox h}$58$^{\mbox m}$17$^{\mbox s}.$217, 
~$-$~$\!\!$42$^{\circ}$52$^{\arcmin}$07$\arcsec.$47, 
agrees with the peak position derived from the SMA observations at 1.3~mm \citep{fra09}.

Assuming that the emission at 880~$\micron$ corresponds to optically thin thermal dust emission, a gas-to-dust ratio of 100, 
a dust temperature of 100~K \citep{guz14,san14}, a dust mass opacity of $\kappa_{880\micron}$=1.8~cm$^{2}$g$^{-1}$ \citep{oss94}, 
and that this object is located at 2.9~kpc, we estimate total H$_{2}$ mass, $M$(H$_{2}$), 
of 15~$\MO$ for core-A and 12~$\MO$ for core-B.
We stress that both cores have similar masses because core-B is defined over a range with an area of four times larger than that of core-A.

\subsection{Spectral line emission}

Figure \ref{fig1} shows the spectra of the $^{12}$CO(3--2) and CH$_{3}$OH(7--6) lines 
integrated within a 23$\arcsec$$\times$23$\arcsec$ region centered on the continuum source. 
The $^{12}$CO(3--2) spectrun shows ``wings" at high velocities that can be attributed to the molecular outflows. 
In fact, we found multiple outflows in the integrated intensity maps (see Sections \ref{out1}, \ref{out2}, and Figures \ref{fig2}(b) and \ref{fig4}).
A broad absorption line, ranging from $\sim-$36 to $-$25~km~s$^{-1}$ with a minimum at $-$29 km~s$^{-1}$, and a few 
narrow absorption lines with the deepest one centered at $-$21.6~km~s$^{-1}$ are also seen in the $^{12}$CO(3--2) spectrum.
Narrow absorption lines might be caused by the cold cloud at foreground, and broad absorption line might be generated 
by combination of spatial filtering (e.g., resolve out) and absorption. 
The CH$_{3}$OH spectrum exhibits emission in a narrower velocity range than that of the $^{12}$CO(3--2) spectrum.

\subsubsection{Identification of two bipolar molecular outflows: $^{12}$CO(3--2) line emission}\label{out1}

From Atacama Pathfinder Experiment (APEX) observations (angular resolution $\sim$ 20\arcsec), \cite{gar07} identified 
a large scale $^{12}$CO(3--2) bipolar outflow, with lobes roughly aligned 
in the north-south direction. The blue-shifted lobe, velocity range from $-$60 to 
$-$38~km~s$^{-1}$, extends up to $\sim$50$\arcsec$ to the south from the 
jet source while the red-shifted lobe, velocity range from $-$22 to $-$0.8~km~s$^{-1}$, extends up to $\sim$50$\arcsec$ to the north. 
The spectrum of the spatially integrated $^{12}$CO(3--2) line emission 
observed with ALMA is quite similar to the APEX $^{12}$CO(3--2) spectrum
centered on the jet source, and both show broad absorption lines.
The broad absorption line feature is most likely produced by the colder, collapsing envelope surrounding the central sources.

Figure \ref{fig2}(b) shows $^{12}$CO(3--2) integrated intensity maps of the blue-shifted 
(velocity range from $-$70 to $-$45.8~km~s$^{-1}$) and 
red-shifted (velocity range from $-$15.9 to $+$8.8~km~s$^{-1}$) wing emission.
The dust continuum emission is overlaid in black contours.  
From Figure \ref{fig2}(b), we identified two bipolar outflows, one aligned in 
the northeast-southwest direction (Outflow-1, R1-B1 in Figure \ref{fig2}(b)), 
and the other aligned in the northwest-southeast direction (Outflow-2, R2-B2 in Figure \ref{fig2}(b)). 
The velocity ranges were selected to separate distinct features of both outflows lobes.
These outflows appear at considerably smaller angular scales 
than the wide-angle bipolar outflow detected with APEX and their position angles are different.

We found another red-shifted component associated with the S1 H$_{2}$O maser cluster to the south of the continuum peak. 
For S1, \cite{rod08} suggested the existence of a YSO candidate associated with this component from their VLA observations. 
This YSO may have a different systemic velocity from that of the central continuum source, 
thus the $^{12}$CO(3--2) component appeared to be dominantly red-shifted with reference to the systemic velocity of the central object.
In fact, we found a faint blue-shifted component, with a velocity range from $-$55 to $-$32~km~s$^{-1}$, associated with S1.
Our observations also support the existence of another source as suggested by \cite{rod08}.
For the discussion, we also plot the positions of the H$_{2}$O (g1 and g2 in Figure \ref{fig2}(a)) 
and CH$_{3}$OH maser clusters (pink crosses in Figure \ref{fig4}) detected by \cite{fra09} and \cite{vor06}, respectively.

\subsubsection{Shock-enhanced regions: CH$_{3}$OH(7--6) line emission}

There are several transitions of CH$_{3}$OH between 338.3~GHz and 338.7~GHz
and many of them were detected with ALMA. For the discussion here, we selected the 
strongest line corresponding to $v_{t}$=0, $J_{K}$=7$_{0}$--6$_{0}$, A-type transition at 338.409~GHz 
\citep{kri10}, hereafter CH$_{3}$OH(7--6).

Figure \ref{fig3}(a) shows the velocity integrated intensity map of the 
CH$_{3}$OH(7--6) emission (color and contours). The CH$_{3}$OH(7--6) emission 
from the central region is strong and well correlated with the dust continuum 
emission. At fainter levels, the CH$_{3}$OH(7--6) emission appears as 
weak filaments elongated to the northeast, southeast, northwest, 
and southwest, forming an hourglass structure. In order to investigate the detailed 
velocity structure of the CH$_{3}$OH(7--6) emission, we produced first and second 
moment maps (Figures \ref{fig3}(b) and \ref{fig3}(c), respectively). 
In the first moment map, we see towards the central region 
a similar velocity gradient as is seen in SO$_{2}$, which might be Keplerian rotation \citep{fra09}.
In the second moment map, the region associated with dust 
continuum emission exhibits a large velocity dispersion of $\sim$10~km~s$^{-1}$. 

CH$_{3}$OH is predominantly formed on grain surfaces \citep{wat02,fuc09}. 
Although small CH$_{3}$OH abundances can be found in cold, quiescent environments \citep{garr07,san13}, 
CH$_{3}$OH is largely released into the gas phase by processes related to active star-formation: 
heating from protostars or sputtering of the grain mantles produced by the interaction of molecular outflows and the ambient gas 
(e.g., Sakai et al. 2013; Yanagida et al. 2014). 
To compare the CH$_{3}$OH(7--6) emission with that of the outflows, we show the second moment map of CH$_{3}$OH(7--6) emission
overlaid with the velocity integrated maps of the blue-shifted and red-shifted $^{12}$CO(3--2) emission in Figure \ref{fig4}.

We find that the northeast and southwest walls of the CH$_{3}$OH(7--6) hourglass structure 
are located at the edge of the northeast and southwest lobes of Outflow-1, 
respectively, suggesting a relationship. We also find that the red-shifted component of Outflow-2, 
R1 is associated with the ring-like wall located southeast of the 
central source. In addition, there is a good correlation between the spatial 
distribution of the CH$_{3}$OH(7--6) emission and the CH$_{3}$OH maser spots.
From all these results, we conclude that the CH$_{3}$OH(7--6) emission traces
zones of post shocked gas produced by the interaction of multiple outflows with the ambient gas within the IRAS\,16547$-$4247 region.

\section{DISCUSSION}
\subsection{Maser distribution and $^{12}$CO(3--2) Outflows}\label{out2}

H$_{2}$O masers \citep{fra09} and class I CH$_{3}$OH maser clusters \citep{vor06}  
have been detected towards the IRAS\,16547$-$4247 region (see Figures \ref{fig2}(a) 
and \ref{fig4}). In high-mass star-forming regions, H$_{2}$O masers are predominantly 
located within a few milliarcseconds from the central driving sources, 
tracing gas motions in the vicinity of the protostar \citep{tor11, chi14}. 
On the other hand, class I CH$_{3}$OH maser clusters are usually excited at the 
shocked interface between a high-velocity jet, or outflow, and the ambient 
gas cloud. The g1 and g2 H$_{2}$O masers detected by \cite{fra09} are located within the continuum peak and 
close to the centers of the two bipolar outflows (see Figure \ref{fig2}(a) and (b)), 
suggesting that they could mark the position of the driving sources of the outflows. 
Higher angular resolution observations are required to identify the driving 
sources and make a more precise relation with g1 and g2. In addition, the H$_{2}$O masers located at the 
position of S1 are most likely excited by a different YSO, thus adding to the multiplicity of the protostars 
in the region.

Figure \ref{fig4} indicates the spatial distribution of the four class I CH$_{3}$OH maser clusters, 
\cite{vor06} present within the field of the ALMA observations.
There are four class I CH$_{3}$OH maser clusters in Figure \ref{fig4}.
The clusters located northeast and southwest of the central source 
roughly aligns with Outflow-1, while the maser cluster seen towards 
the northwest aligns with the blue-shifted lobe of Outflow-2, B2.
The dashed lines in Figure \ref{fig4} indicate the directions of the outflows estimated by using the position of CH$_{3}$OH maser clusters.
The northern cluster coincides with the location of the the jet knots of \cite{rod08}, 
and may have been excited by the shock influence of the jet. A schematic diagram to visualize 
the star formation activities in the region based on previous observations and our ALMA results is shown in Figure \ref{fig5}.

\subsection{Nature of the IRAS\,16547$-$4247 region}

In an attempt to explain the observed morphology of the shocks traced by the 
CH$_{3}$OH(7--6) emission, we discuss two possible scenarios that may be responsible 
for the hourglass structure seen in Figures \ref{fig3} and \ref{fig4}.

The first of the plausible scenarios is the precession of the biconical jet reported 
by \cite{rod08}(see Figure \ref{fig4}). The interaction of the high-velocity precessing jet with the 
dense ambient medium results in strong shocks, which will trace a larger biconical 
structure (as seen in the morphology of the narrow-angle CH$_{3}$OH(7--6) wall). 
Assuming that the opening angle of the observed narrow-angle CH$_{3}$OH(7--6) wall is the angle subtended by the minor axis 
at a distance of one-half the major axis, we derived the 
opening angle of the narrow-angle CH$_{3}$OH(7--6) wall to be $\sim$45~$^{\circ}$. 
The jet has an opening angle of 15$^{\circ}$, implying that the jet will need to precess through 30$^{\circ}$ 
to create the observed narrow-angle CH$_{3}$OH(7--6) wall.
\cite{rod08} assumed that the jet seems to be precessing linearly with time, and derived a precession rate, $\beta$, of 0$^{\circ}$.08~yr$^{-1}$ 
($\beta$=$\alpha$$v$, with $\alpha$$\sim$2$^{\circ}$.3~arcsec$^{-1}$ and a jet velocity, $v$, of $\sim$490~km~s$^{-1}$ as in Rodr{\'{\i}}guez et al. 2008).
Assuming that these rates of the N1 and S1 represent the precession rate of the entire jet, 
we estimated that it would take the collimated precessing jet about $\sim$~380~yr to produce the observed morphology.

\begin{figure}
\epsscale{0.7}
\plotone{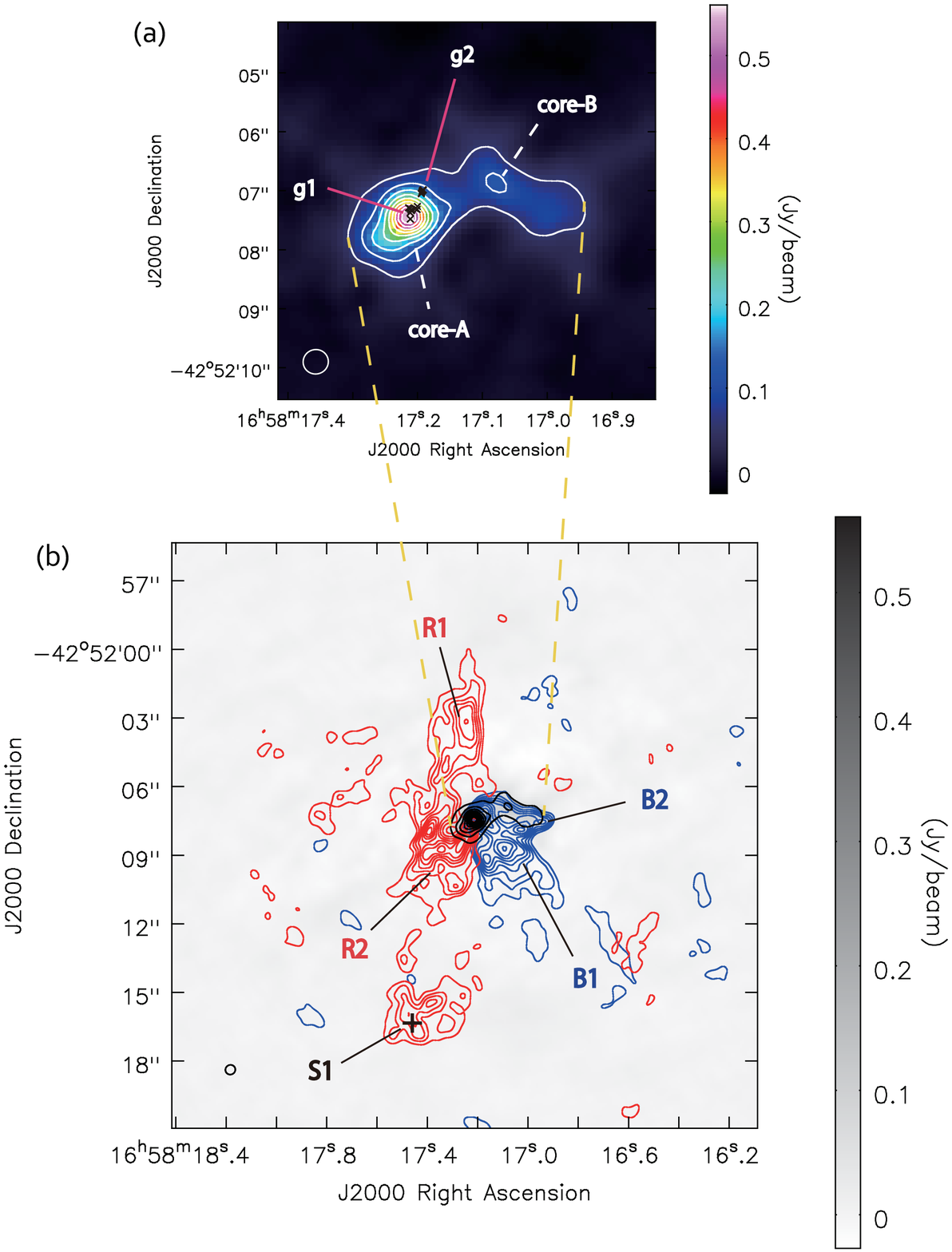}
\caption{(a) ALMA 880~$\micron$ continuum emission (color and contours) from IRAS\,16547$-$4247.
The small crosses (black) mark the positions of the water maser clusters named g1 and g2 in \cite{fra09}.
The white contours range from 10 to 90~$\%$ of the peak emission, in steps of 10~$\%$.
(b) Integrated intensity map of the $^{12}$CO(3--2) outflows and the continuum emission (black contours). 
The blue color represents blue-shifted gas, while the red color represents the red-shifted gas. 
The integrated velocity range of $^{12}$CO(3--2) for the blue-shifted side is from $-$70.0 to $-$45.8~km~s$^{-1}$, 
while for the red-shifted side is from $-$15.9 to $+$8.8~km~s$^{-1}$.
The contours for integrated intensity maps, with intervals of 3$\sigma$, start from the 3$\sigma$ level
(1$\sigma$=0.6~Jy~beam$^{-1}$~km~s$^{-1}$ for both blue-shifted and red-shifted gas).
The black cross marks the position of the water masers named S1 in \cite{fra09}.
The gray-scale bar shows the flux density of dust continuum emission.}
\label{fig2}
\end{figure}

\begin{figure}
\epsscale{1}
\plotone{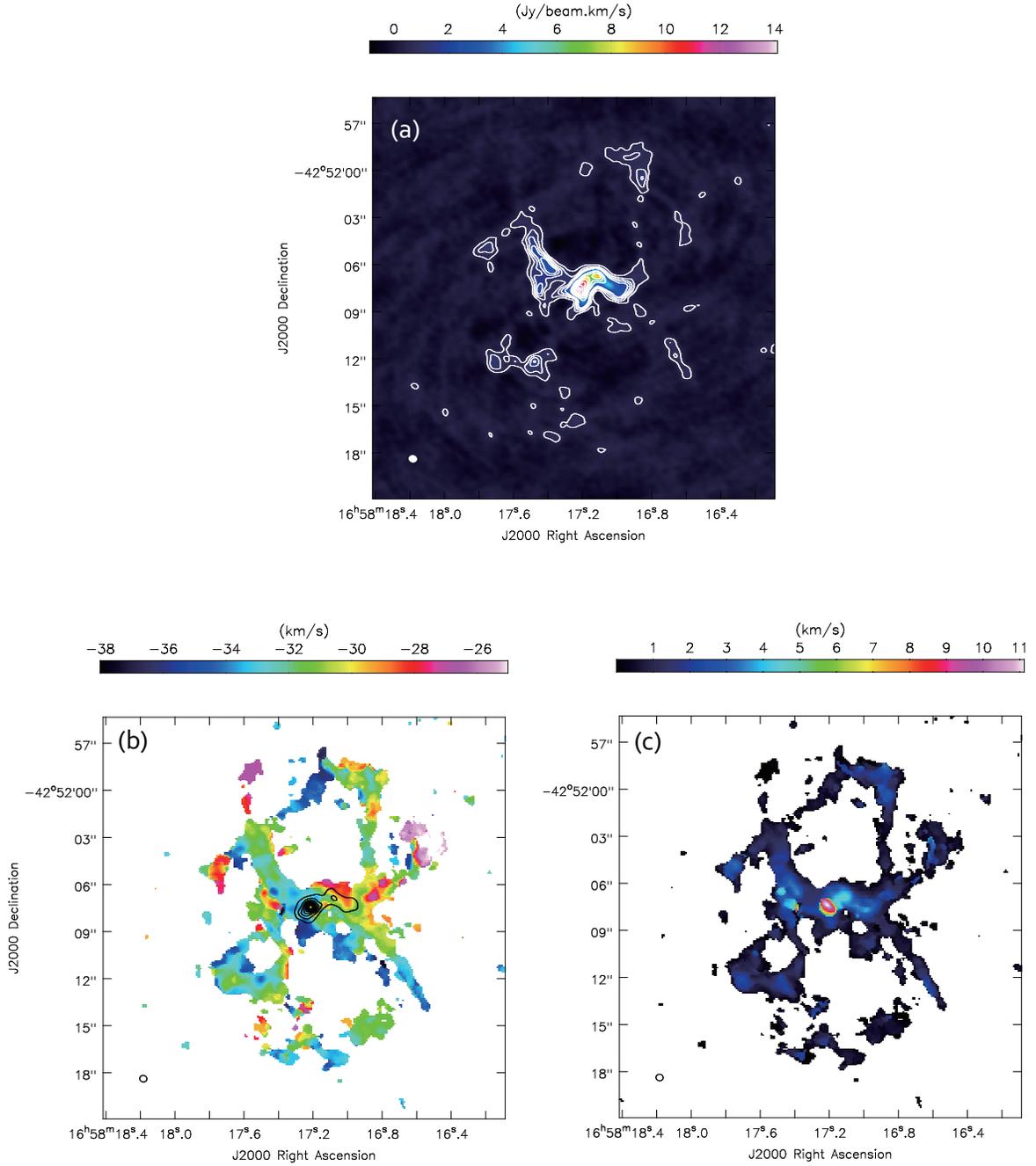}
\caption{
(a): Integrated intensity map of the CH$_{3}$OH(7--6) emission (color and contours).
Contours start first from 3$\sigma$ to 30$\sigma$ increasing in intervals of 3$\sigma$ levels and
then they continue in step of 10$\sigma$ up to the 100$\sigma$ level (1$\sigma$=0.15~Jy~beam$^{-1}$~km~s$^{-1}$).
(b): Dust continuum emission (black contours) superposed on the first moment map of CH$_{3}$OH(7--6) emission (color). 
(c): The second moment map of CH$_{3}$OH(7--6) emission (color). }
\label{fig3}
\end{figure}

\begin{figure}
\epsscale{0.8}
\plotone{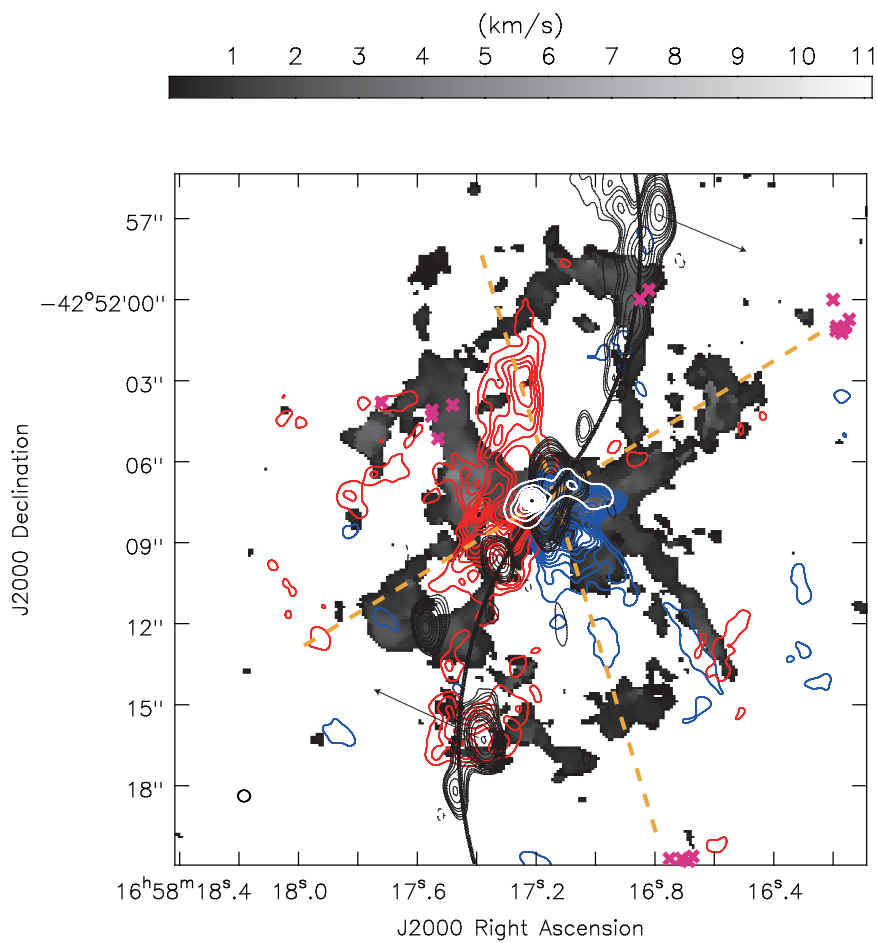}
\caption{The second moment map of the CH$_{3}$OH(7--6) emission (gray scale) and integrated intensity map of the $^{12}$CO(3--2) outflows (contours).
The integrated velocity range for the blue-shifted and red-shifted side are the same in the Figure \ref{fig2}(b).
The contours with the intervals of the 3$\sigma$ levels start from the 3$\sigma$ levels (see Figure \ref{fig2}(b)).
White contours show the dust continuum emission as in Figure \ref{fig2}(a).
Black contours show the VLA continuum emission detected by \cite{rod08}.
The pink crosses mark the positions of the class I CH$_{3}$OH masers listed in \cite{vor06}.
The dashed lines indicate the direction of outflows estimated by using the position of maser clusters.}
\label{fig4}
\end{figure}

\begin{figure}
\rotate
\epsscale{0.6}
\plotone{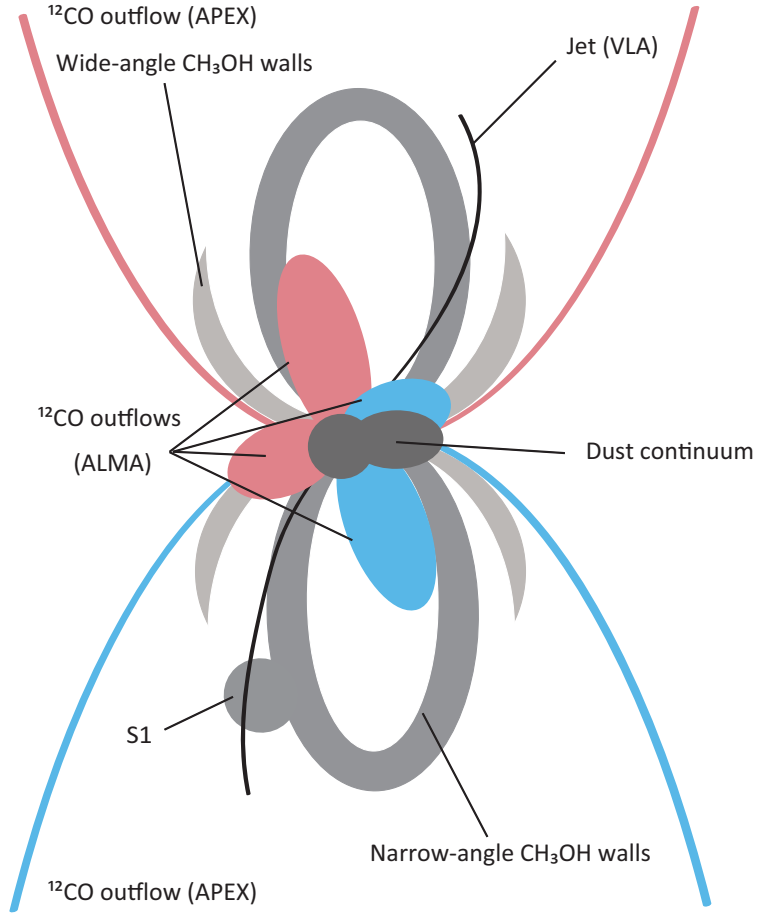}
\caption{A schematic picture of the complex environment revealed by ALMA in IRAS\,16547$-$4247.
The dust continuum cores are surrounded by interacting outflows, cavity walls, and an ionized jet.}
\label{fig5}
\end{figure}

The second scenario is that of multiple, or at least two, driving sources powering intersecting outflows. 
\cite{gar03} showed two outflow lobes, which may be suggestive of a single driving source. 
But $^{12}$CO(3--2) at high-angular resolution with ALMA shows two outflows; one aligned in northwest-southeast direction, 
and the other one aligned in the northeast-southwest direction (Figures \ref{fig2}(b) and \ref{fig4}). 
This points to the multiplicity of the protostars driving outflows in the region. 
This agrees with the distribution of the class I CH$_{3}$OH masers of \cite{vor06}, which also suggest the presence of more 
than one outflow source in the region. The complexity of the outflow structures could be explained by interaction of the 
outflowing materials from different driving sources.

\bigskip
We thank the anonymous referee for constructive comments that helped to improve this manuscript.
We also thank the ALMA staff for the observations during the commissioning stage. 
G.G. acknowledges support from CONICYT project PFB-06.
This letter makes use of the following ALMA data: ADS/JAO.ALMA 2011.0.00419.S. 
ALMA is a partnership of ESO (representing its member states), NSF (USA), and NINS (Japan), 
together with NRC (Canada), NSC, and ASIAA (Taiwan), in cooperation with the Republic of Chile. 
The Joint ALMA Observatory is operated by ESO, AUI/NRAO, and NAOJ.
Data analysis were carried out on common use data analysis computer system at the Astronomy Data Center, 
ADC, of the National Astronomical Observatory of Japan.
Finally, we acknowledge Masao Saito for his contribution to our study.

\appendix

\begin{thebibliography}{}

\bibitem[Chibueze et al.(2014)]{chi14} Chibueze, J.~O., Omodaka, T., Handa, T., et al.\ 2014, \apj, 784, 114 

\bibitem[Eisl{\"o}ffel et al.(2000)]{eis00} Eisl{\"o}ffel, J., Smith, M.~D., \& Davis, C.~J.\ 2000, \aap, 359, 1147 

\bibitem[Fuchs et al.(2009)]{fuc09} Fuchs, G.~W., Cuppen, H.~M., Ioppolo, S., et al.\ 2009, \aap, 505, 629

\bibitem[Garay et al.(2007)]{gar07} Garay, G., Mardones, D., Bronfman, L., et al.\ 2007, \aap, 463, 217 

\bibitem[Garay et al.(2003)]{gar03} Garay, G., Brooks, K.~J., 
Mardones, D., \& Norris, R.~P.\ 2003, \apj, 587, 739 

\bibitem[Garrod et al.(2007)]{garr07} Garrod, R.~T., Wakelam, V., \& Herbst, E.\ 2007, \aap, 467, 1103 

\bibitem[Guzm{\'a}n et al.(2014)]{guz14} Guzm{\'a}n, A.~E., Garay, G., Rodr{\'{\i}}guez, L.~F., et al.\ 2014, arXiv:1410.0233 

\bibitem[Franco-Hern{\'a}ndez et al.(2009)]{fra09} Franco-Hern{\'a}ndez, R., Moran, J.~M., Rodr{\'{\i}}guez, L.~F., 
\& Garay, G.\ 2009, \apj, 701, 974 

\bibitem[Hills et al.(2010)]{hil10} Hills, R.~E., Kurz, 
R.~J., \& Peck, A.~B.\ 2010, \procspie, 7733,  

\bibitem[Higuchi et al.(2014)]{hig14} Higuchi, A.~E., 
Chibueze, J.~O., Habe, A., Takahira, K., \& Takano, S.\ 2014, \aj, 147, 141 

\bibitem[Higuchi et al.(2013)]{hig13} Higuchi, A.~E., Kurono, 
Y., Naoi, T., et al.\ 2013, \apj, 765, 101 

\bibitem[Higuchi et al.(2010)]{hig10} Higuchi, A.~E., Kurono, 
Y., Saito, M., \& Kawabe, R.\ 2010, \apj, 719, 1813 

\bibitem[Higuchi et al.(2009)]{hig09} Higuchi, A.~E., Kurono, 
Y., Saito, M., \& Kawabe, R.\ 2009, \apj, 705, 468 


\bibitem[Kristensen et al.(2010)]{kri10} Kristensen, L.~E., van Dishoeck, E.~F., van Kempen, T.~A., et al.\ 2010, \aap, 516, AA57 

\bibitem[Lada(2010)]{lad10} Lada, C.~J.\ 2010, Royal Society 
of London Philosophical Transactions Series A, 368, 713 

\bibitem[Lada 
\& Lada(2003)]{lad03} Lada, C.~J., \& Lada, E.~A.\ 2003, \araa, 41, 57 

\bibitem[McKee 
\& Ostriker(2007)]{mck07} McKee, C.~F., \& Ostriker, E.~C.\ 2007, \araa, 45, 565 

\bibitem[Ossenkopf \& Henning(1994)]{oss94} Ossenkopf, V., \& Henning, T.\ 1994, \aap, 291, 943 

\bibitem[Ridge et al.(2003)]{rid03} Ridge, N.~A., Wilson, 
T.~L., Megeath, S.~T., Allen, L.~E., \& Myers, P.~C.\ 2003, \aj, 126, 286 

\bibitem[Rodr{\'{\i}}guez et al.(2005)]{rod05} 
Rodr{\'{\i}}guez, L.~F., Garay, G., Brooks, K.~J., 
\& Mardones, D.\ 2005, \apj, 626, 953 

\bibitem[Rodr{\'{\i}}guez et al.(2008)]{rod08} 
Rodr{\'{\i}}guez, L.~F., Moran, J.~M., Franco-Hern{\'a}ndez, R., et al.\ 
2008, \aj, 135, 2370 

\bibitem[Sakai et al.(2013)]{sak13} Sakai, T., Sakai, N., 
Foster, J.~B., et al.\ 2013, \apjl, 775, L31 

\bibitem[S{\'a}nchez-Monge et 
al.(2014)]{san14} S{\'a}nchez-Monge, {\'A}., Beltr{\'a}n, M.~T., Cesaroni, R., et al.\ 2014, \aap, 569, AA11 

\bibitem[Sanhueza et al.(2013)]{san13} Sanhueza, P., Jackson, J.~M., Foster, J.~B., et al.\ 2013, \apj, 773, 123 

\bibitem[Torrelles et al.(2011)]{tor11} Torrelles, J.~M., 
Patel, N.~A., Curiel, S., et al.\ 2011, \mnras, 410, 627 

\bibitem[Yanagida et al.(2014)]{yan14} Yanagida, T., Sakai, 
T., Hirota, T., et al.\ 2014, \apjl, 794, L10 

\bibitem[Voronkov et al.(2006)]{vor06} Voronkov, M.~A., 
Brooks, K.~J., Sobolev, A.~M., et al.\ 2006, \mnras, 373, 411 

\bibitem[Watanabe \& Kouchi(2002)]{wat02} Watanabe, N., \& Kouchi, A.\ 2002, \apjl, 571, L173 

\bibitem[Zinnecker 
\& Yorke(2007)]{zin07} Zinnecker, H., \& Yorke, H.~W.\ 2007, \araa, 45, 481 

\end{thebibliography}
\end{document}